\documentclass[iop,apj,tighten,twocolappendix,numberedappendix]{emulateapj}
\usepackage{apjfonts}

\usepackage{graphicx}
\usepackage{amsmath}
\usepackage{amssymb}
\usepackage{color}
\usepackage{mathtools}
\usepackage[breaklinks,colorlinks,citecolor=blue,linkcolor=red]{hyperref} 
\usepackage[all]{hypcap}

\newcommand\msun{\, \rm M_\odot}

\newcommand\kms{\, \rm km\,s^{-1}}

\begin{document}

\title{Tidal disruption events, main-sequence extreme-mass ratio inspirals and binary star disruptions in galactic nuclei}

\author{Re'em Sari\altaffilmark{1} and Giacomo Fragione\altaffilmark{1}}
\affil{$^1$Racah Institute for Physics, The Hebrew University, Jerusalem 91904, Israel}
\thanks{E-mail: sari@phys.huji.ac.il; giacomo.fragione@mail.huji.ac.il}

\begin{abstract}
The Galactic Center has been under intense scrutiny in the recent years thanks to the unprecedented missions aiming at measuring the gas and star dynamics near the supermassive black hole (SMBH) and at finding gravitational wave (GW) signatures of inspiralling stellar black holes. In the crowded environment of galactic nuclei, the two-body interactions alter the distribution of stars on long timescales, making them drift in energy and angular momentum. We present a simplified analytical treatment of the scattering processes in galactic stellar nuclei, assuming all stars have the same mass. We have discussed how the interplay between two-body relaxation and gravitational wave emission modifies the slope of the inner stellar cusp within the SMBH sphere of influence, and calculated the rates of tidal disruption events (TDEs) and main-sequence extreme-mass ratio inspirals (MS-EMRIs) of stars that are tidally disrupted by the SMBH. We find that typically the ratio of the TDE and MS-EMRI rates is the square of the ratio of the tidal and Schwarzschild radii. For our Galaxy, this implies that the rate of MS-EMRIs is just about a percent of the TDE rate. We then consider the role of stars injected on highly eccentric orbits in the vicinity of the SMBH due to Hills binary disruption mechanism, and show that the MS-EMRI rate can almost approach the TDE rate if the binary fraction at the SMBH influence radius is close to unity. Finally, we discuss that physical stellar collisions affect a large area of phase space.
\end{abstract}


\section{Introduction}

Most of the galaxies over the whole Hubble sequence harbour super massive black holes (SMBHs), with masses in the range $10^6\ \mathrm{M}_{\odot}\lesssim M \lesssim 10^{10}\ \mathrm{M}_{\odot}$, in their innermost regions \citep{kor13}. Dense and complex structures of stars, stellar remnants and gas surround SMBHs (see \citet{alex17} for a recent comprehensive review). The Milky Way's Galactic Center (GC) is the only nucleus close enough to resolve scales smaller than $\sim$ pc \citep{gil17,baumg18,gall18}. The recent big advance in dedicated instruments, e.g. GRAVITY\footnote{https://www.eso.org/sci/facilities/paranal/instruments/gravity.html} \citep{eise11,grav2018a,grav2018b}, allows to observe with an unprecedented precision the GC, which marks an unique opportunity of improving the understanding of our Galaxy and galactic nuclei in general.

In the GC, stars and compact remnants move in the smooth near-Keplerian potential of the SMBH, which dominates the dynamics within the radius of influence $R_h$, beyond which the potential of the SMBH is overcome by the galactic field \citep{mer13}. On timescales much longer than the orbital period, the microscopic fluctuations of the potential make stars energy and angular momentum diffuse, as a result of continuous non-coherent scatterings with other stars. Stars are subject to a net residual specific force $\propto \sqrt{N}$ ($N$ is the number of stars), and their energy and angular momentum diffuse on the typical two-body timescale $T_{\rm 2B}\gtrsim 10^9$-$10^{10}$ yr \citep{baror14,baror16}. As a consequence, a population of equal-mass stars rearrange their orbits and relax into a cuspy density profile $n\propto r^{-7/4}$ \citep[$r$ is the radial distance with respect to the SMBH;][BW]{bahcall76}. Mass spectrum, and the subsequent dynamical friction, affects the typical slope, with more massive objects that relax in a steeper profile \citep{bahcall77,hopale06}. Massive perturbers may play a role as well \citep{perets07}. Only galactic nuclei harbouring SMBHs less massive than $\sim 10^7\ \mathrm{M}_{\odot}$ have typical evolutionary timescales small enough to make the effects of the uncorrelated stellar interactions important within a Hubble time.

On timescales smaller than $T_{\rm 2B}$, but longer than the stars orbital period, the residual torque $\propto \sqrt{N}$ resulting from fluctuations of the average potential becomes relevant in shaping the stars orbits \citep{rauch96}. Both the direction and magnitude of the angular momentum (hence eccentricity) diffuse on a resonant relaxation timescale, $T_{\rm RR}\gtrsim 10^7$-$10^{9}$ yr. On even smaller timescales, only the transverse component of the residual torque has a non-negligible effect, thus shaping the inclinations of the orbital planes of the stars on a vector resonant relaxation timescale, $T_{\rm VRR}\gtrsim 10^5$-$10^{7}$ yr \citep{kocs11,kocs15}.

In addition to star-star interactions, the gravitational wave (GW) emission radiation can dissipate energy. When energy is efficiently dissipated in the innermost regions of the cusp, stars and stellar black holes gradually inspiral and illuminate the GW sky as an EMRI \citep[extreme-mass ratio inspiral;][]{hop06,aharon16}. In the last orbits, relativistic precession decouples the GW inspiral from the residual torques of the background stars, thus resulting in a slow inspiral that roughly conserves the pericenter of the orbit \citep{pet64}, while shrinking its semimajor axis by losing energy at each pericenter passage, until the object is swallowed by the SMBH. This object can either be a stellar black hole \citep[BH-EMRI;][]{hopmale05} or a main-sequence star \citep[MS-EMRI;][]{linial2017}. Characterizing BH-EMRIs and MS-EMRI is of extremely interest for the upcoming Laser Interferometer Space Antenna (LISA)\footnote{https://lisa.nasa.gov/}, which is expected to probe SMBH demographics in galactic nuclei and cosmological parameters even at large redshifts \citep{lisa17}. 

While in the GW regime stars are disrupted after gradually losing energy, stars can be swallowed by directly plunging onto the SMBH. Unlike the GW mergers, the plunge disruption scenario requires the star to remain on a plunging orbit only long enough to pass through its periapsis once, where the star is disrupted and shines as a tidal disruption event (TDE) \citep{stone13}. The typical radius within which a star is disrupted is the tidal disruption radius $R_T \approx R_*(M/m)^{1/3}$, where $R_*$ and $m$ are the radius and mass of the star, respectively. Stars are driven onto plunging loss-cone orbits by the continues two-body scattering events within the SMBH sphere of influence, which randomizes the stars energy and angular momentum. However, the angular momentum evolves much faster than energy for eccentric orbits, thus rendering the angular momentum diffusion as the driving mechanism for producing TDEs \citep{alex17}.

Both observations and theory suggest that stars can be transported to the innermost part of the cusp of stars by binary star disruptions \citep{hills88,brw14,brw15}. In the inner regions of galactic nuclei, binary stars undergo three-body exchange interactions with the SMBH, where one of the stars is ejected as hypervelocity star (HVS) with velocities of hundreds km s$^{-1}$, while the former companion remains bound to the SMBH \citep{yut03,sari10,kobay12,ross14}. Recently, \citet{kop2019} reported the observation of a HVS with a velocity of $1755\pm 50\kms$ in the Galactic frame. Triple and quadruples stars may undergo the same fate as well \citep{fgu18,fgi18}. Other mechanisms have been proposed to explain the observed population of HVSs in our Galaxy \citep{yut03,cap15,fra16,frack17}, but binary disruptions remain the favored scenario. However, recent analyses show a growing evidence of high-velocity objects possibly not originated in the GC \citep{boub2018,marchrb2018,delaf2019}. Little attention has been devoted to the long term effects of binary disruptions in the ecology of galactic nuclei. \citet{mill05} showed that the tidal breakup of black hole binaries can produce events observable with low eccentricity in the LISA band. \citet{bromley2012} examined the fate of the stars that remain bound to the SMBH after the binary disruption, and found that most of them undergo TDEs, thus fuelling the growth of the SMBH. Recently, \citet{frasar18} have shown that continuous injection of stars enabled by binary disruptions may make the cusp slope steeper, whose extent and importance depend on the injection rate and survival fraction of the injected stars.

In this paper, we provide a simplified analytical treatment of the scattering processes in galactic stellar nuclei, assuming all stars have the same mass. This problems contains two dimensionless numbers. First, the mass ratio $M/m$, which is also the number $N_h$ of stars within the SMBH influence radius. Second, the ratio between $R_h$ and the SMBH Schwarzchild radius $R_s=2GM/c^2$. By coincidence, these dimensionless numbers are equal in our Galaxy, $M/m\sim R_h/R_s \sim 4\times 10^6$. More massive galaxies tend to have $N_h>R_h/R_s$, and vice versa. We also discuss the role of collisions, and focus on the effect of binary disruptions on the rate of TDEs and MS-EMRIs. 

The paper is organized as follows. In Section \ref{sect:timescales}, we discuss the typical two-body relaxation and GW timescales, and how they shape the cusp of stars surrounding the SMBH. In Section \ref{sect:binaries}, we explain the role of binary disruptions, which is then investigated by means of numerical simulations in Section \ref{sect:simulations}. Finally, in Section \ref{sect:conclusions}, we discuss the implications of our findings and draw our conclusions.

\section{Two-body relaxation and gravitational waves}
\label{sect:timescales}

Consider a stellar cusp, with $N(r)$ stars, each of mass $m$ and with semimajor axis $r$ around a SMBH of mass $M$. This is related to the number density of stars by $N(r)\approx n(r) r^3$. The two-body relaxation time at radius $r$ is given by \citep{bahcall76}
\begin{equation}
T_{2B}(r)=\frac{P(r)}{\ln \Lambda}\left(M \over m\right)^2 {1 \over N(r)}\ ,
\end{equation}
where $P(r)$ is the star's orbital period and $\ln \Lambda\sim 10$ is the Coulomb logarithm. This is the timescale by which a star at a roughly circular orbit changes it angular momentum and energy by a factor of order unity.

For stars on very eccentric orbits, the orbital angular momentum $J$ is much smaller than the circular angular momentum at the same semimajor axis $J_c$, and the timescale to significantly change their angular momentum is smaller by $(J/J_c)^2$. If we denote their periapsis distance by $r_p \approx (J/J_c)^2 r$, then we can define the eccentric orbit relaxation time as \citep{binney87}
\begin{equation}
T^{J}_{2B}(r,r_p)=\frac{P(r)}{\ln \Lambda}\left(M \over m\right)^2 {1 \over N(r)} \left( r_p \over r \right)\ .
\label{eqn:2bj}
\end{equation}
For very eccentric orbits with $r_p\ll r$, this is the typical timescale to change the periapsis distance of the star's orbit. In the previous expression, we assumed that angular momentum changes are dominated by interactions at $r$. This holds as long as $N(r)r$ is an increasing function of $r$, so as long as the density profile is shallower than $n(r) \propto r^{-4}$, which we will assume is the case.

Very eccentric stars also relax their energy faster than stars on circular orbits of the same semimajor axis. This is because the kicks they get from the inner denser cluster are more significant to their energy change. The cross section for some energy change $\Delta E$ is given by $(Gm/\Delta E)^2$, independent of the velocities of the stars. As a consequence, the inner cusp will dominate the scatterings as long as $n(r) r$ is a decreasing function of $r$, which we will assume to be true. The typical timescale to change the semimajor axis for an eccentric star is therefore given by \citep{binney87}
\begin{equation}
T^{E}_{2B}(r,r_p)=\frac{P(r)}{\ln \Lambda}\left(M \over m\right)^2 {1 \over N(r_p)} \left( r_p \over r \right)^2=\left(r \over r_p \right)^{1/4} T^{J}_{2B}\ .
\end{equation}
Energy relaxation for highly eccentric orbits is therefore slightly less efficient than that of angular momentum, if the BW density profile ($n\propto r^{-7/4}$) is assumed. Density profiles steeper than $\rho(r)\propto r^{-2}$ will result in energy relaxation being faster than that of angular momentum. 

\begin{figure}
\centering
\includegraphics[scale=0.575]{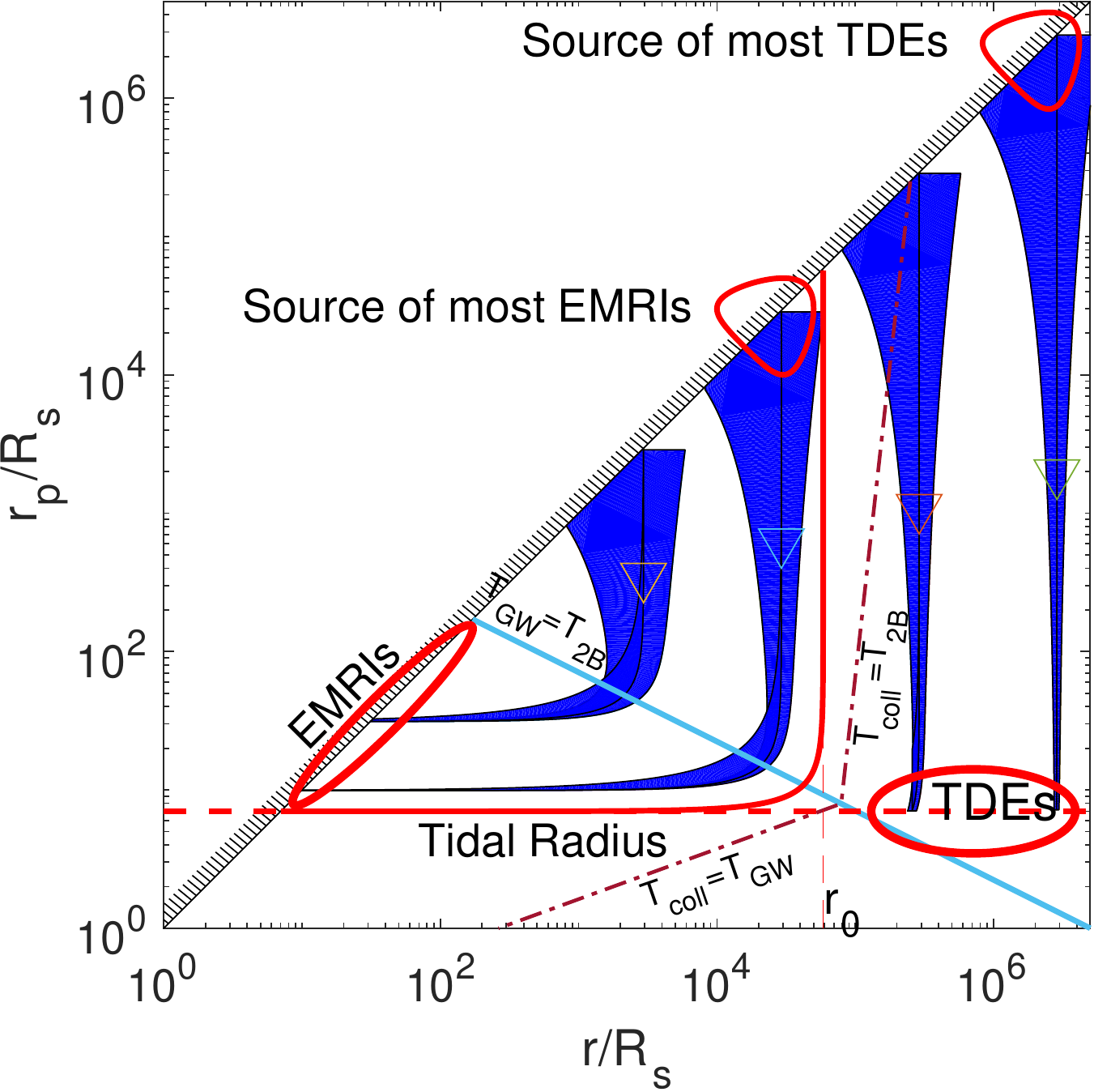}
\caption{Stellar orbital evolution by two-body encounter, gravitational waves and breakup of binaries. TDEs are typically generated by stars coming from roughly the influence radius on an extremely eccentric orbit, over its dynamical time. Most EMRI-stars revolving around the SMBH on a slowly shrinking circular orbit have evolved from orbits with initial semimajor axis $r_0\sim R_h(R_s/R_T)^2$. Our analysis ignores collisions between stars, though for main sequence objects, these collisions are important in a large fraction of the parameter space, bounded by the thin dash-dot line, given by Eq.~\ref{eq:col2b} and Eq.~\ref{eq:colgw}.}
\label{fig:tdeemris}
\end{figure}

As discussed, GWs circularize the orbit of highly eccentric orbits, while keeping the periapsis distance roughly fixed. This happens on a timescale \citep{pet64,hopale06}
\begin{equation}
T_{GW}(r,r_p)={R_s \over c} {M \over m} \left( r_p \over R_s \right)^4 \left(r \over r_p \right)^{1/2}\ ,
\label{eqn:gw}
\end{equation}
where $R_s=2GM/c^2$ is the SMBH Schwarzchild radius.

Equating the timescales of GWs (Eq.~\ref{eqn:gw}) to that of periapsis evolution by two-body encounters (Eq.~\ref{eqn:2bj})
\begin{equation}
\left(R_s \over r_p \right)^{5/2}\left(M \over m\right) {1 \over \ln \Lambda\ N(r)}=1\ .
\end{equation}
In the case of a BW cusp profile, $N(r)=(M/m)(r/R_h)^{5/4}$, hence
\begin{equation}
\left. r_p \over R_s \right. =(\ln \Lambda)^{-2/5}\left(r \over R_h\right)^{-1/2}\ .
\label{eqn:equal}
\end{equation}
Therefore, a star at the radius of influence with pariapse close to the Schwarzschild radius has equal time to shrink its semimajor axis by GWs and to change its periapsis by scatterings. In the $r_p$-$r$ plane, stars which are above the line given by Eq.~\ref{eqn:equal}, evolve primarily due to two-body scatterings, while stars below it mostly shrink in semimajor axis by emission of GWs. This is given by line $T_{GW}=T_{2B}$ in Fig.~\ref{fig:tdeemris}. It is remarkable that this line passes through the point $(r=R_h,r_p=R_s)$ regardless of the two dimensionless parameters in the problem, $M/m$ and $R_h/R_s$ (neglecting the factor $(\ln \Lambda)^{-2/5}$).

We note that the above calculation assumes that it takes more than one orbit to change the stars' semimajor axis or periapsis significantly. On the line given by Eq.~(\ref{eqn:equal}), the number of orbits in such a timescale is (assuming an BW profile)
\begin{equation}
\mathcal{N}=(\ln \Lambda)^{-7/5} \left(M \over m\right) \left( R_s \over R_h \right) \left( r \over R_h \right)^{-11/4}\ .
\end{equation}
For our Galaxy, the first product is of order unity (by coincidence), so it takes about a single orbit for a star at the radius of influence and $r_p=R_s$ to change it periapsis or apoapsis significantly. However, anywhere within the radius of influence, $\mathcal{N}\gg 1$, thus our previous assumption is satisfied for our Galaxy. Using the $M \propto \sigma^4$ relation \citep{mer01}, this is even more readily satisfied for any larger galaxy, but only marginally at the radius of influence of smaller galaxies. Note that $\mathcal{N}>1$ is also the condition that a loss cone defined by the line (\ref{eqn:equal}) is empty. Usually, the loss cone is defined by the tidal radius or the Schwarzschild radius. Yet, it makes more sense to treat Eq.~\ref{eqn:equal} as defining the loss cone, as for smaller angular momenta gravitational waves deplete the stars and provide an effective zero boundary condition for the diffusion the occurs for larger angular momenta by two body interactions.

As mentioned, there are two important dimensionless parameters in this problem, $\left(M/m\right)$ and $\left(R_h /R_s\right)$. The relation between these parameters is important. The above discussion shows that $\left(M/m\right) \gg \left(R_h /R_s\right)$ results in an empty loss cone at the radius of influence, and with a loss cone defined by the Schwarzschild radius. For our Galaxy, by coincidence, both are $\left(M/m\right)\approx\left(R_h /R_s\right)\approx 4\times 10^6$, then it is a borderline situation. However, as we discuss later, the tidal radius in our Galaxy for main sequence stars is larger than the event horizon, and the loss cone associated with the tidal radius is therefore empty at the radius of influence.

The line of equal timescale cuts that of a circular orbit, i.e. $r=r_p$, for 
\begin{equation}
r=(\ln \Lambda)^{-4/15} R_s^{2/3}R_h^{1/3}\ .
\end{equation}
For our Galaxy, this is $\sim 100$ times the Schwarzschild radius. Below this distance, the BW cusp is modified and GWs shrinks the orbit. Interestingly, a constant flux of stars through circular orbits, would imply $N(r) \propto r^4$, i.e. a decreasing density towards the center. However, the flux of stars on circular orbits decaying by GWs is not constant as it is supplemented by the flux of starts evolving first by two-body encounters and then circularizing by GWs (along the curving dark cones of Fig.~\ref{fig:tdeemris}). Therefore, instead, we demand that the flux of stars through circular orbits of size $r< R_s^{2/3}R_h^{1/3}$ \footnote{We are neglecting the factor $(\ln \Lambda)^{-4/15}\sim 0.6$.} would equal the flux of stars supplied by two-body encounter into $r_p=r$, i.e. from initial circular orbits of size $r_i=R_h (r/R_s)^{-2}$. The flux of these is 
\begin{eqnarray}
\frac{N(r)}{T_{GW}(r)} 
=\frac{1}{ P(R_h)}\left(r\over R_s\right)^{-2}\ .
\end{eqnarray}
This implies 
\begin{equation}
N(r) = \left(r\over R_s\right)^{2} \left(R_h \over R_s\right)^{-3/2} {M \over m} \propto r^2\ .
\end{equation}
The overall cusp is therefore given by
\begin{equation}
N(r)={M \over m}    
\begin{cases}
{\left(r\over R_s\right)^2
\left(R_h \over R_s\right)^{-3/2}
} & { {\rm for\,}  r<R_s^{2/3}R_h^{1/3}  } \cr
\cr
\left( r \over R_h \right)^{5/4} & {\rm for\,} r>R_s^{2/3}R_h^{1/3}
\end{cases}
\end{equation}
For our Galaxy, this implies a single star below a distance of $\sim 50 R_s$. Yet, this point is within the regime of GWs, which act on each star separately. The low number of stars, even when it falls below unity, does not invalidate our solution and represents the expectation value for the number of stars.

It is easy to generalize this equation to calculate also the number of eccentric orbits, i.e., instead of $N(r)$, we could estimate $N(r,r_p)$ - the number of stars with semimajor axis smaller than $r$ and periapsis smaller than $r_p$. This is given by
\begin{equation}
N(r,r_p)={M \over m}    
\begin{cases}
{\left(r_p\over R_s\right)^{3/2}\left(r\over R_s\right)^{1/2}
\left(R_h \over R_s\right)^{-3/2}
} & { {\rm for\,}  r_p<R_s (r/R_h)^{-1/2}  } \cr
\cr
\left( r \over R_h \right)^{1/4}\left( r_p \over R_h \right) & {\rm for\,} r_p>R_s (r/R_h)^{-1/2}  
\end{cases}
\end{equation}
Now, $N(r,r_p)=1$ is given by 
\begin{equation}
{r_p\over R_s}=    \left(m \over M\right)^{2/3} \left(R_h \over R_s \right) \left( r \over R_s \right)^{-1/3}\ .
\end{equation}
Accidentally, for the parameters of our Galaxy, there is a single object with semimajor axis equal to the radius of influence, but periapsis as small as the Schwarzschild radius. This is another consequence of the coincidence $\left(M/m\right) \approx \left(R_h /R_s\right)$.

\subsection{Tidal disruption events and main-sequence extreme-mass ratio inspirals}
\label{subsec:tdesemris}

For small enough galaxies, tidal disruption of main sequence stars occurs before the star hits the Schwarzschild radius \citep{stone13}. If this encounter happens on an extremely eccentric orbit, this result in a violent TDE of the star, over its dynamical time. If, instead, the star approaches on a slowly shrinking circular orbit, the star transfers mass to the SMBH on the GW timescale \citep{dai2013,linial2017}. These events are know as main sequence MS-EMRIs. \citet{linial2017} showed that for main-sequence stars mass transfer may result is an expanding orbit causing an inverted \textit{"Chirp"} signal which (spelled backwards) we call a \textit{"Prich"}, where the GW frequency decreases with time. These events may also appear as a sequence of TDE-like flares, if two (or more) consecutive MS-EMRIs collide \citep{metz2017}. Figure~\ref{fig:tdeemris} illustrates the different origins of TDE and MS-EMRI events.

In this picture, the rate of TDEs is dominated by the supply from roughly circular orbits at the radius of influence $R_h$
\begin{equation}
\mathcal{R}_{TDEs}=\frac{N(R_h)}{\ln(J_c/J_{LC}) T_{2B}(R_h)}\approx \frac{1}{P(R_h)}\ ,
\end{equation}
simply the inverse period at the radius of influence. In the previous equation, $J_c$ and $J_{LC}=\sqrt{2GM R_T}$ are the circular and loss-cone angular momentum, respectively. On the other hand, the rate of MS-EMRIs is given by the supply from roughly circular orbits of the largest radius that does not result in a TDE. The periapsis of these orbits, once their evolution is dominated by GWs, is just above the tidal radius $R_T$. Their initial radius is therefore 
\begin{equation}
r_0=R_h \left(R_s \over R_T \right)^2\ ,
\end{equation}
and the rate of these events is given by 
\begin{equation}
\mathcal{R}_{MS-EMRIs}\approx\frac{N(r_0)}{T_{2B}(r_0)}=\frac{1}{P(R_h)} \left(R_s \over R_T \right)^{2}\ .
\end{equation}
The ratio between the rate of MS-EMRIs and that of TDEs is simply
\begin{equation}
{ \mathcal{R}_{MS-EMRIs} \over \mathcal{R}_{TDEs} } \approx \left( R_s \over R_T \right)^2\ .
\label{eqn:ratiorert}
\end{equation}
For our Galaxy, and Solar-like stars, $R_T \sim 10 R_s$, so the MS-EMRI rate is about a percent of the TDE rate.

\subsection{Collisions}

Close enough to the SMBH, the orbital velocities become larger than the typical escape velocity from the surface of a star ($v_{esc}\sim 600 \kms$ for $1\msun$ star), and collisions become more likely than scattering events. Whether it results in a merger or a destruction, depends on the mass ratio of the colliding stars, the ratio between their relative velocity to their surface escape speed as well as the impact parameter of the collision (grazing or head on) \citep{benz1987,trac2007,gabu2010}. This lead \cite{alex17} to define the collision radius where the velocity dispersion equals the escape velocity. He estimated the collision rate, the inverse collision time, as
\begin{equation}
T_{\rm coll}^{-1}={N(r) \over r^3} \left(\frac{GM}{r}\right)^{1/2}R_*^2\left[1+\left(\frac{Gm}{R_*}\frac{r}{GM}\right)^2\right]\ .
\end{equation}
where the term in parenthesis is the gravitational focusing term, increasing the effective cross section for collisions beyond the physical size of the star. However, this term is significantly larger than unity only if the velocity dispersion of the stars is smaller than their escape speed and in that case the collision time is longer than the relaxation time. We can therefore ignore gravitational focusing when collisions are important.

However, as we have shown, the interesting objects observationally, those that lead to tidal disruption events or those that start evolve quickly due to gravitational waves, are extremely eccentric. The collisional time is not determined only by their semimajor axis, but also depends on their periapsis distance. We therefore have to generalize the result of \cite{alex17} to highly eccentric orbits. Collisions around pericenter would dominate as long as the optical depth, $N(r)/r^2$, decreases with radius. Such is the case for a Bahcall-Wolf cusp or any density profile with $\alpha>1$.
We therefore use the eccentric collisional time as
\begin{equation}
T_{\rm coll}^{-1}={N(r_p) \over r_p^2 r} \left(\frac{GM}{r}\right)^{1/2}R_*^2.
\label{eqn:tcoll}
\end{equation}

Equating $T_{\rm coll}$ to $T_{2B}^J$ (Eq.~\ref{eqn:2bj}), we find
\begin{equation}
r_p=\left[ R_* \left(M \over m \right) \ln \Lambda^{-1/2} \right]^{-8} r^9.
\label{eq:col2b}
\end{equation}
Note that the quantity in square brackets is the collisional radius result of \cite{alex17}.

For stars on small enough periapsis and semimajor axis GW rather than two body scattering dominate the evolution. Equating
the collision time to $T_{GW}$ (Eq.~\ref{eqn:gw}) we obtain:
\begin{equation}
r_p=\left( M \over m \right)^{-8/11} \left( R_h \over R_s \right)^{5/11} \left( R_* \over R_s \right)^{8/11} \left(r \over R_s \right)^{4/11} R_s
\label{eq:colgw}
\end{equation}
We show the phase space where collisions are important in Fig.~\ref{fig:tdeemris}. We do not take into account collisions in our calculations, but it is clear that they would affect a significant portion of the phase space. A realistic description of galactic nuclei must take collisions into account. We leave this to future work.

\section{The effect of binaries}
\label{sect:binaries}

Binary disruption results in one star ejected, perhaps as a HVS, while the other remains tightly bound to the SMBH \citet{hills88}. \citet{frasar18} showed that binary breakup is an important source of stars close to the SMBH, and may modify the cusp structure if binaries are sufficiently abundant.

These stars are injected with periapsis distance ($r_{p,inj}$) comparable to that of the binary tidal radius, and a semimajor axis ($r_{inj}$) larger than that by a factor of $\sim (M/m)^{1/3}\approx 100$, almost independently of the original semimajor axis (as long as it is larger than that)
\begin{eqnarray}
r_{p,inj}&=&\alpha R_* \left(M \over m \right)^{1/3}=\alpha R_T\\
r_{inj}&=&\alpha R_* \left(M \over m \right)^{2/3}=\alpha R_T \left( M \over m \right)^{1/3}\ .
\label{eqn:rpinjrinj}
\end{eqnarray}
In the previous equations, $\alpha$ is the initial binary semimajor axis $a_b$ in terms of the stellar radius. The stars injected into the tightest orbits are those who broke from an almost contact binary, i.e. $\alpha$ is of the order of a few.

The criterion for these injected stars to be mostly dominated by GW evolution is (see Eq.~\ref{eqn:equal})
\begin{equation}
\frac{r_{p,inj}}{R_s}<(\ln \Lambda)^{-2/5} \left( r_{inj} \over R_h \right)^{-1/2}\ .
\end{equation}
We now define a critical binary separation parameter $\alpha_c$ above which the captured member of the binary is mostly affected by two-body scatterings
\begin{equation}
\alpha<\alpha_c=  (\ln \Lambda)^{-4/15} {R_s \over R_T}  \left( R_h \over R_s \right) ^{1/3}  \left(m \over M \right)^{1/9} 
\end{equation}
The binaries with orbital separation parameter of $\alpha_c$ dissolve on a periapsis distance of
\begin{equation}
{r_{p,c} \over R_s}= (\ln \Lambda)^{-4/15} \left( R_h \over R_s \right) ^{1/3}  \left(m \over M \right)^{1/9}\ ,  
\end{equation}
while their semimajor axis is larger by a factor of $(M/m)^{1/3}$
\begin{equation}
{r_{c} \over R_s}= (\ln \Lambda)^{-4/15} \left( R_h \over R_s \right) ^{1/3}  \left(M \over m \right)^{2/9} .
\end{equation}
For the parameters of the Milky Way and Solar-mass main sequence stars, $R_T/R_s\approx 10$, hence $\alpha_c \approx 3$.

\begin{figure}
\centering
\includegraphics[scale=0.575]{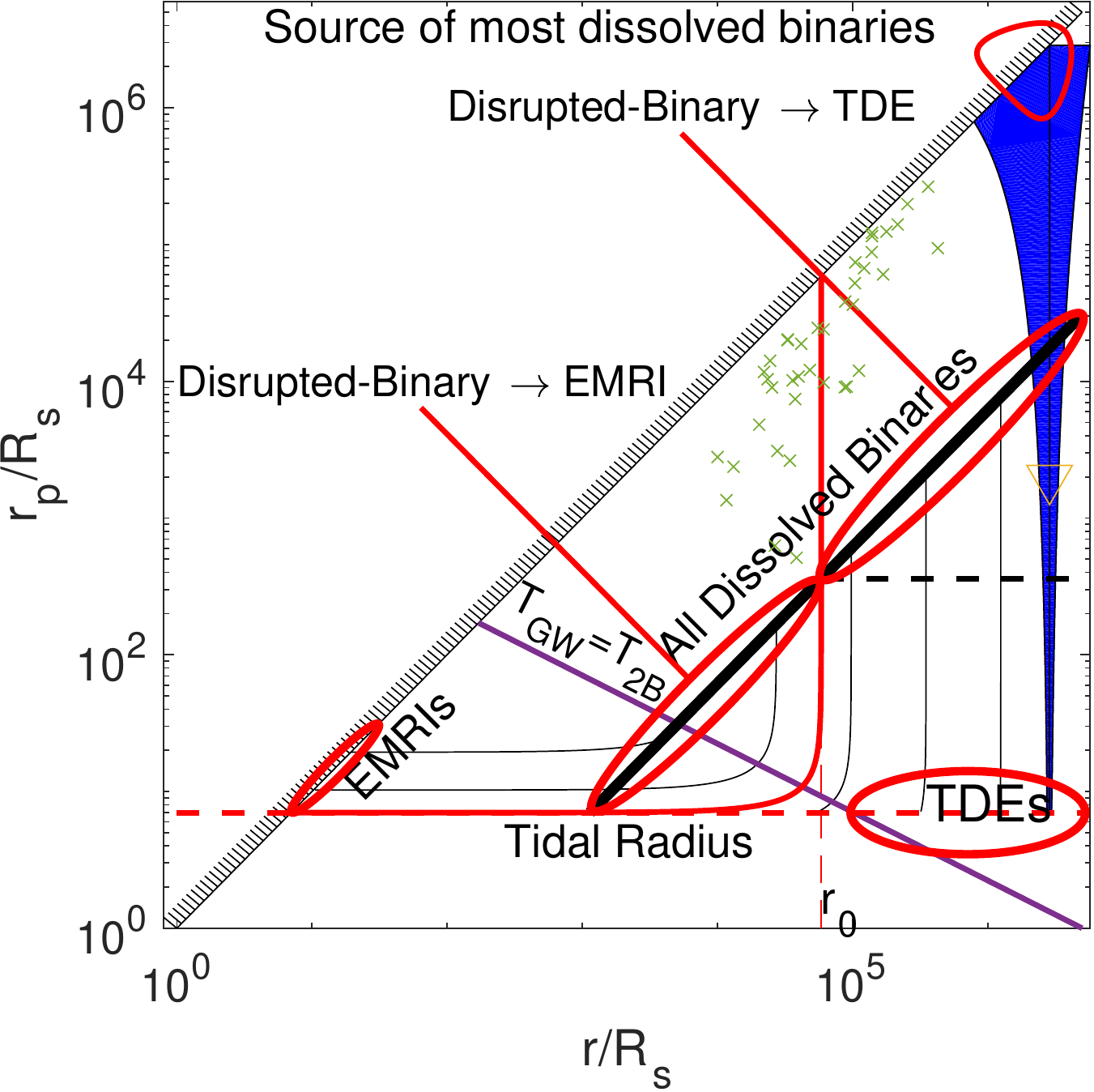}
\caption{Influence of the binary disruption channel on the rate of TDEs and MS-EMRIs. Disrupted binaries have typically semimajor axis from a few hundreds to a few thousands AU \citep{frasar18}. Green crosses represent the actual measured orbits of S-stars \citet{gil17}. The solid thin line represent the likely evolution of the captured binary member by two body interactions and gravitational waves. The thick red line separate the region that can lead to MS-EMRIs. The region of EMRIs shown here is smaller than that of Figure \ref{fig:tdeemris} as it depicts only those that could evolved from the captured member of a dissolved binary.}
\label{fig:binaries}
\end{figure}

If the fraction of disrupted binaries is large enough, the injected stars modify the cusp profile. This increases the two-body encounter rate, and shortens the two-body relaxation time, somewhat reducing the value of $\alpha_c$. \citet{frasar18} showed that the cusp obtains a shape of
\begin{equation}
N(r)={M \over m} \times 
\begin{cases}
\left(\eta R_h \over r_{min} \right)^{1/2} \left(r \over R_h\right)^{5/4} &\ {\rm for\,}\ r<r_{min} \\
\\
\eta^{1/2} \left(r \over R_h \right)^{3/4}  &\ {\rm for\,}\ r_{min}<r<\eta R_h \cr
\\
\left(r \over R_h \right)^{5/4} &\ {\rm for\,}\ \eta R_h<r
\end{cases}
\label{eqn:cuspbinaries}
\end{equation}
where $\eta$ is the fraction of binaries at the influence radius $R_h$, and $r_{min}$ is the minimum injected radius, which depends on the minimum separation of binaries. The outer section of the cusp, $\eta R_h<r<R_h$, is unaffected by the binaries. We also note that the extent of the modified cusp also depends on the fraction of the injected stars that survive without being dissolved either as TDEs or due to GW emission. \citet{frasar18} estimated that $\sim 70$-$80$\% of the injected stars diffuse to TDE orbits. 

However, once we include the influence of GWs, initially tight binaries that got dissolved can be circularized by two-body-scatterings into orbits smaller than $r_{p,c}$. We can derive this further population by requiring that the flux of stars $N(r)/T_{GW}$ equals the rate of injected from binary disruptions. This adds an additional population of stars on very tight orbits in the following amount:
\begin{equation}
N(r)=\left(M \over m \right)^{5/9}\left(R_s \over R_h \right)^{1/6}\eta_B \times 
\begin{cases}
\left (r \over r_{p,c} \right)^4 & r<r_{p,c} \\
\left(\frac{r}{r_{p,c}}\right)^{1/2} & r_{p,c}<r
\end{cases}
\label{eqn:circbin}
\end{equation}
In the above equation, $\eta_B=\eta/\ln(a_{\max}/a_{\min})$ is the fraction of tight binaries
per logarithmic unit of separation. Assuming the fraction of binaries is $\eta=0.1$ and a log-uniform binary separation distribution (see Sect. 4.1) in the range $a_{\rm min}=0.01$\,AU-$a_{\rm max}=1$\,AU for solar-mass stars, then $\eta_B\sim 0.04$. For the Milky Way, the coefficient in Eq.~\ref{eqn:circbin} for this extra-component is $400\eta_B\approx 17$. As a consequence, we expect a population of $\sim 20$ Solar-mass stars in the GC, orbiting at $r_{p,c}\approx 30 R_S\approx 3{\rm AU}$ at orbital periods below a day.

\section{Numerical simulations}
\label{sect:simulations}

Direct $N$-body simulations would represent the ideal tool to investigate the evolution and distribution of stars in the proximity of a SMBH. Unfortunately, prohibitive computational times limit $N$-body simulations to a small number of stars, of the order of $\sim 50$k stars, thus far from the $\sim 10^6$ stars that reside in the SMBH sphere of influence \citep{bau04a,bau04b,bau17}.

In what follows, we describe the computational method we adopt to perform long-term evolution of the stellar cusp around the Milky Way's SMBH. Compared to \citet{frasar18}, we upgraded the scheme by including angular momentum evolution and GW energy loss. To summarize, our method has four main features
\begin{itemize}
\item It follows the 2-D evolution of the energy and angular momentum (or semimajor axis and eccentricity, respectively) of each star, therefore conserving the number of stars;
\item The rate of scattering changes with energy and angular momentum, and with the number of particles that have similar energy and angular momentum;
\item It takes into account relativistic GW effects;
\item It takes into account possible source terms, e.g. from breaking of binaries.
\end{itemize}
We focus our attention on the region inside the sphere of influence, where the stellar dynamics is dominates by the SMBH field
\begin{equation}
R_h=\frac{GM}{\sigma^2}\approx 2\ \mathrm{pc}\ ,
\end{equation}
where $\sigma$ is the velocity dispersion external to the radius of influence. In all our calculations, we assume $M=4\times 10^6\ \mathrm{M}_{\odot}$ \citep{gil17} and a single-mass population of stars of mass $m=1\ \mathrm{M}_{\odot}$. The region of interest spans a wide range of energy, angular momentum and distances with respect to the SMBH. The innermost radius we take into account is the tidal disruption radius of $1\ \mathrm{M}_{\odot}$ star \citep{stone13}
\begin{equation}
r_{in}= R_*\left(\frac{M}{m}\right)^{1/3}\approx 1\ \mathrm{AU}\ .
\end{equation}

The semimajor axis and eccentricity of the stars are continuous variables, but we divide the $(r;r_p)$ space into bins, thus considering a 2D grid. Every time step we count how many stars $N(i,j)$ ($i$ is the index for the semimajor axis, $r_i$, and $j$ is the index for the pericenter, $r_{p,j}$) are in a given square of the grid, and how many stars are in a given semimajor axis bin, regardless of the pericenter
\begin{equation}
N(i)=\sum_j N(i,j)\ .
\end{equation}
To determine the timestep, we compute the angular momentum two-body timescale for each square of the grid (see Eq.~\ref{eqn:2bj})
\begin{equation}
T_{2B}^{J}(i,j)=\frac{r_{p,j}}{R_h}\sqrt{\frac{r_i}{R_h}} \frac{F}{N(i)}\ ,
\end{equation}
where
\begin{equation}
F=\frac{f_t P(R_h) (M/m)^2}{\gamma \log \Lambda}\ .
\end{equation}
Here, we set $f_t=0.1$, a reduction factor for the timescale, and $\gamma=1.5$, a factor that takes into account the bin size \citep{frasar18}. In analogy to \citet{frasar18}, we choose as overall timestep the minimum of the two-body timescales
\begin{equation}
\Delta T=\min_{i,j} T_{2B}^{J}(i,j)\ ,
\end{equation}
such that the total time at the k-th step is $T_k=T_{k-1}+\Delta T$. Updating at every time step all the stars in each square of the 2D grid is time-consuming. Hence, we check for each square of the 2D grid if
\begin{equation}
T_k\gtrsim L(i,j)+T_{2B}^{J}(i,j)\ ,
\label{eqn:comptk}
\end{equation}
where $L(i,j)$ is the moment when we made the last update (with respect to the overall time) of a given square $(i,j)$ of the grid. If Eq.~\ref{eqn:comptk} is satisfied, we update the energy and angular momentum of the stars in the square $(i,j)$. For each of these stars, we find the different bins explored during its orbital motion and identify $W1$ and $W2$ such that the orbit of the star is entirely within the bins $(r_{W1},r_{W2})$. Then, we compute the average impact parameter $B(W)$ in each of the bins it went through as
\begin{equation}
B(W)=r_W \sqrt{\frac{P_*}{N(W)T_{2B}^{J}(i,j)}}\ ,
\end{equation}
where $P_*$ is the orbital period of the given star. The typical shifts in energy and angular momentum are computed as \begin{eqnarray}
\Delta E(W)&=&\frac{Gm}{B(W)} \\
\Delta J(W)&=&r_W \frac{Gm}{v(W) B(W)}=\frac{r_W}{v(W)}\Delta E(W)\ ,
\end{eqnarray}
where $v(W)=\sqrt{GM/r_W}$ is the velocity of the star in the bin. Finally, we update the energy and the angular momentum of each star in each bin
\begin{eqnarray}
E_{new}&=&E_{old}+\sin{\chi} \Delta E \\
J_{new}&=&\left({J_{old}^2+\Delta J^2-2 J_{old}\Delta J \cos \Phi}\right)^{1/2}\ ,
\end{eqnarray}
where $0\le \chi < 2\pi$ and $0\le \Phi < 2\pi$ are drawn randomly from a uniform distribution.

After updating energy and angular momentum, we check the following conditions
\begin{itemize}
\item if $J_{new}\le J_{LC}=\sqrt{2GM R_T}$, the star is considered a TDE and removed 
\item if $r_{new}=GM/E_{new}\ge R_h$, the star is considered escaped from the cusp and removed
\end{itemize}
In these cases, a new star is generated randomly in the last bin and with eccentricity drawn from a thermal distribution \citep{frasar18}. 

In the case GW effects are taken into account\footnote{We take into account GWs for all the stars with $r_p<0.1 R_h$.}, we update the energy and angular momentum of the stars as \citep{hopale06,hop06}
\begin{eqnarray}
E_{new}&=&E_{old}+\Delta E_{GW} \\
J_{new}&=&J_{old}+\Delta J_{GW}\ ,
\end{eqnarray}
where
\begin{eqnarray}
\Delta E_{GW}&=&\frac{8\pi}{5\sqrt{2}}f(e)\frac{m c^2}{M}\left(\frac{r_p}{r_S}\right)^{-7/2} \frac{T_{2B}^{J}(i,j)}{P(r)} \\
\Delta J_{GW}&=&-\frac{16\pi}{5}g(e)\frac{Gm}{c}\left(\frac{r_p}{r_S}\right)^{-2} \frac{T_{2B}^{J}(i,j)}{P(r)}\ ,
\end{eqnarray}
and
\begin{eqnarray}
f(e)&=&\frac{1+(73/24)e^2+(37/96)e^4}{(1+e)^{7/2}} \\
g(e)&=&\frac{1+(7/8)e^2}{(1+e)^2}\ .
\end{eqnarray}

If binary injection is taken into account, we parametrize the rate of disrupted binaries with the dimensionless parameter $\eta$. Following \citet{frasar18}, we generate in each timestep
\begin{equation}
N_b=\eta\frac{\Delta T}{P(R_h)}
\label{eqn:binj}
\end{equation}
injected stars as a consequence of the binary tidal disruption, that are added to the pre-existing population. We sample injected star semimajor according to $f(a) \propto 1/a$, and we set their eccentricities to $1-(m/M)^{1/3}\approx0.99$ \citep[Eq.~\ref{eqn:rpinjrinj}; see also][]{brw15,alex17}.

\begin{figure} 
\centering
\includegraphics[scale=0.55]{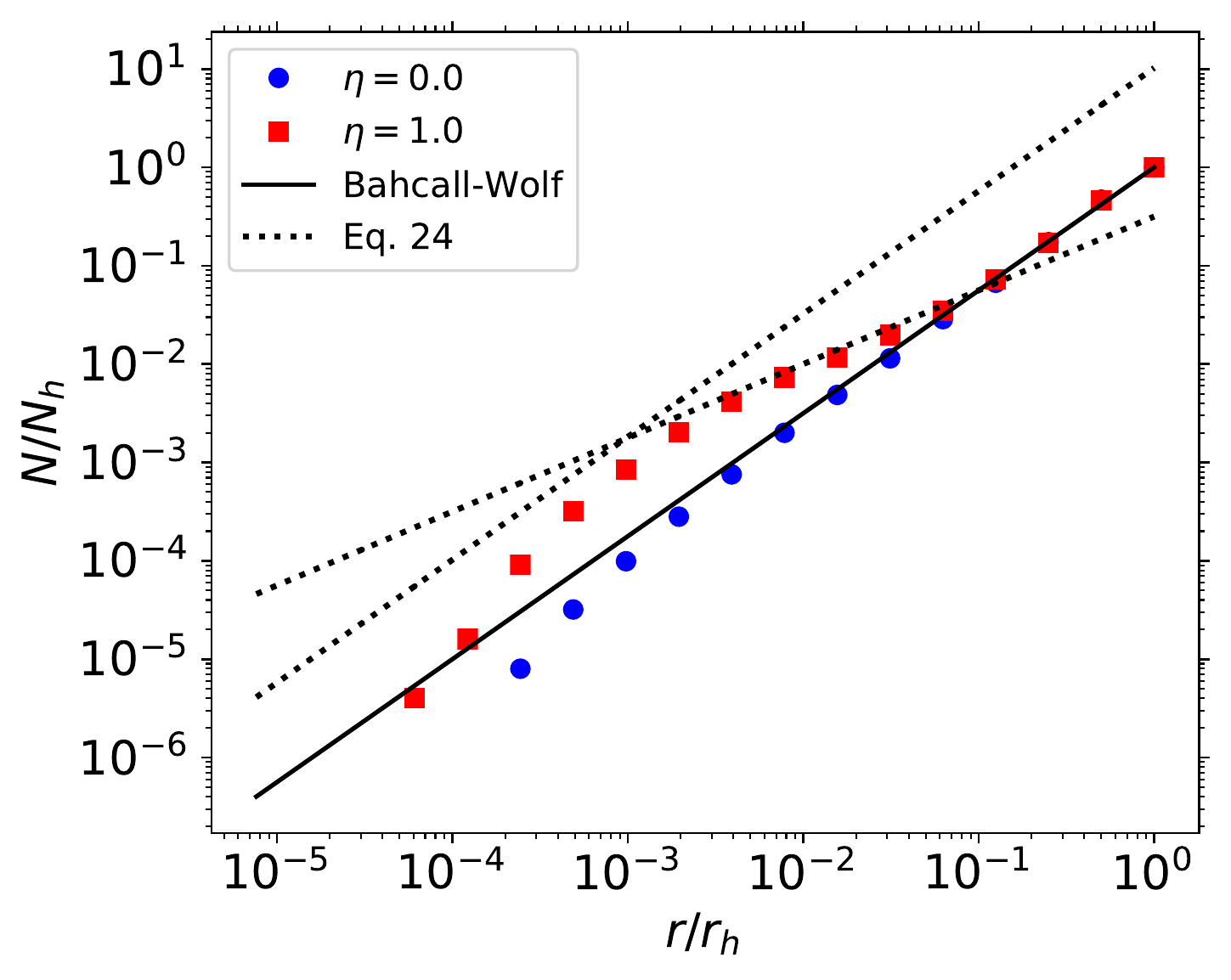}
\caption{Cusp profile with no binaries and with injection of binaries with $\eta=1$. The standard Bahcall-Wolf cusp $N(r)/N_h=(r/R_h)^{5/4}$ (solid curve) and the binary-modified profile from Eq.~\ref{eqn:cuspbinaries} (dotted curves) are shown as reference. In the binary-modified profile we used $\eta_{\rm eff}=0.1\eta=0.1$.}
\label{fig:cuspbin}
\end{figure}

\subsection{Cusp slope, tidal disruption events and main-sequence extreme-mass ratio inspirals}

\begin{figure*} 
\centering
\begin{minipage}{20.5cm}
\includegraphics[scale=0.55]{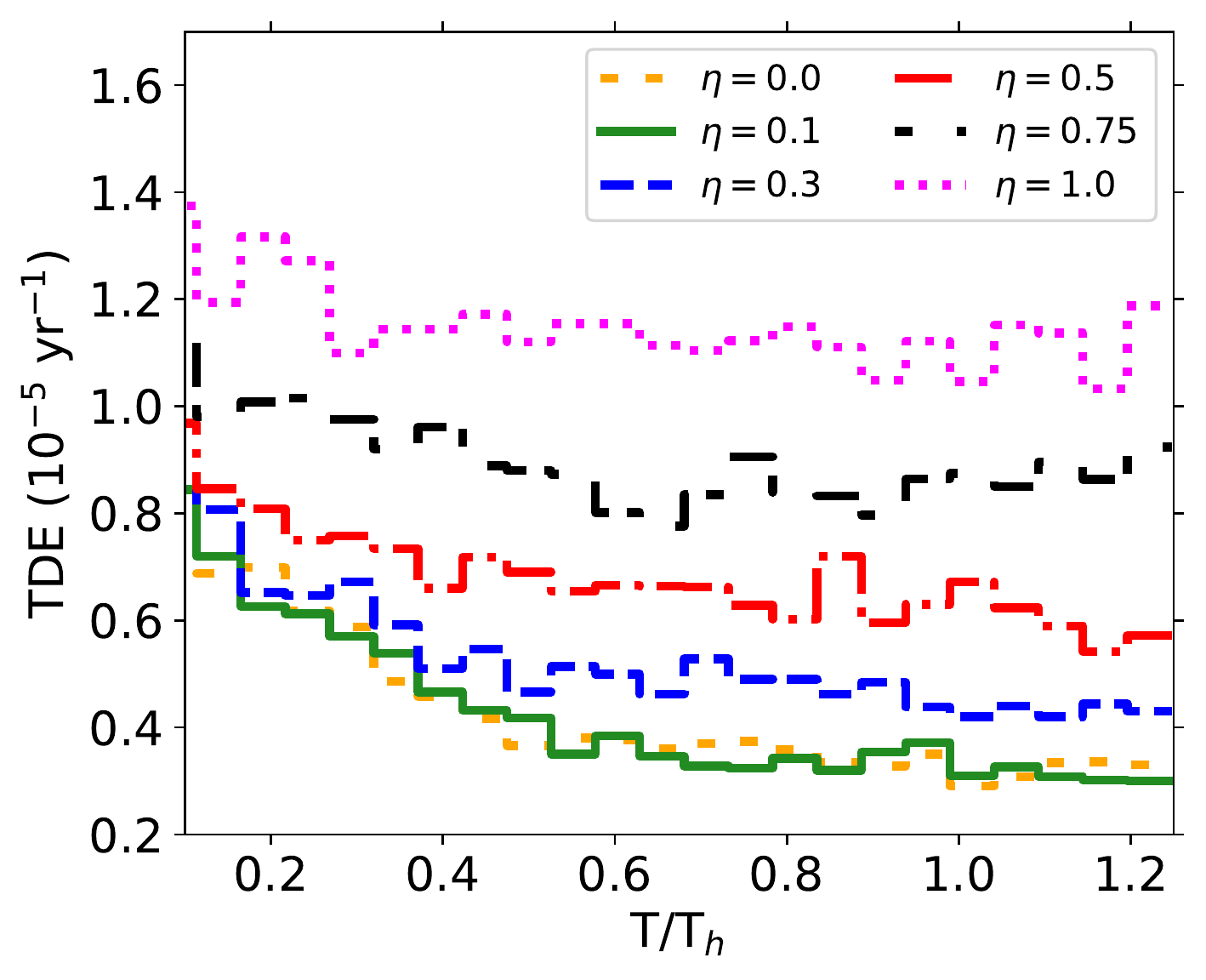}
\hspace{1.5cm}
\includegraphics[scale=0.55]{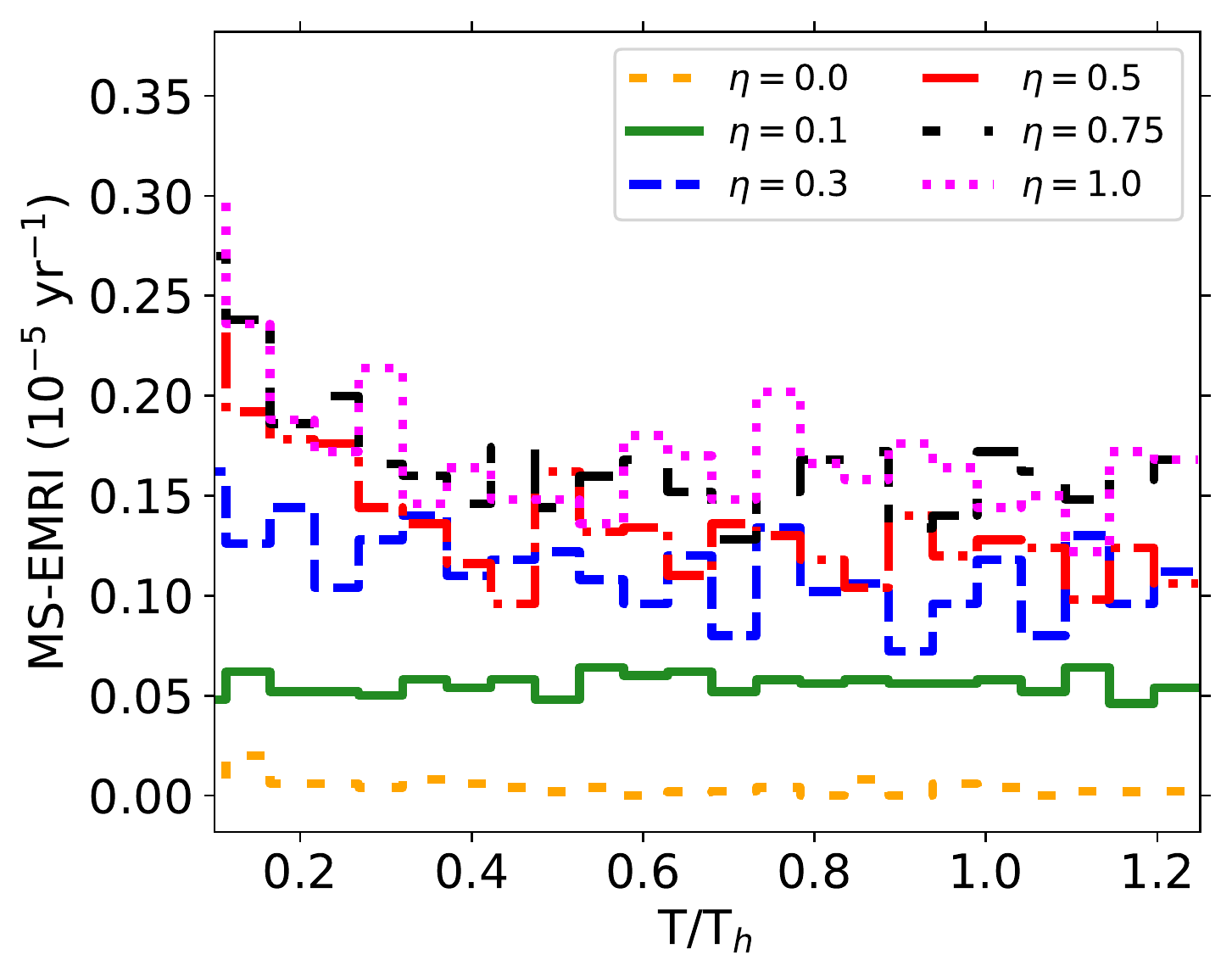}
\end{minipage}
\caption{TDE (left) and MS-EMRI (right) rates as a function of the binary fraction $\eta$. The TDE rate results significantly modified only when $\eta\gtrsim 0.3$. While with no injection of stars from binary disruptions $\Gamma_{MS-EMRI}$ is just a few percent of $\Gamma_{TDE}$, the MS-EMRI rate becomes $\sim 1/5$ of the TDE rate if the injection of binaries is significant ($\eta\gtrsim 0.5$).}
\label{fig:tdeemris_sim}
\end{figure*}

In our simulations, we consider $0\le \eta\le 1$ ($\eta=0$ means no dissolved binaries are injected), and study how the cusp slope, the TDE rate and the MS-EMRI rate change as a function of the binary fraction $\eta$. We consider $10^6$ stars and start with a stable \citet{bahcall76} cusp, i.e star semimajor axis follow a distribution $N(r)\propto r^{5/4}$. For initial momenta, we sample uniformly $0\le (J/J_c)^2 \le 1$ ($J_c=\sqrt{GMr}$ is the circular angular momentum of a given semimajor axis $r$), i.e. thermal distribution of eccentricities. When dissolved binaries are injected, we assume that the binary semimajor axis are distributed according to \citep{duq91}
\begin{equation}
f(a_b)\propto \frac{1}{a_b}
\end{equation}
in the interval $(a_{min}$,$a_{max})$ at the SMBH radius of influence, and set the semimajor axis of the captured stars according to the same distribution scaled by a factor $\sim (M/m)^{2/3}$ \citep{frasar18}
\begin{equation}
f(r)\propto\frac{(M/m)^{2/3}}{r}
\label{eqn:frbhbin}
\end{equation}
in the range $r_{min}=(M/m)^{2/3} a_{min}$ and $r_{max}=(M/m)^{2/3} a_{max}$, with eccentricity $e\sim 0.99$. For Solar-mass stars, we chose the minimum as $a_{min}=0.01$ AU, while for the maximum we set $a_{max}=0.1$ AU, since binaries with larger semimajor axis are typically disrupted by the background stars rather than by the SMBH \citep{hop09}.

Figure~\ref{fig:cuspbin} illustrates the profile $N(r)/N_h$ of the simulations along with the theoretical curves $\propto r^{5/4}$ and $\propto r^{3/4}$ from Eq.~\ref{eqn:cuspbinaries}. For $\eta=0$, our results recover the standard \citet{bahcall76} solution, where energy conservation dictates $\alpha=5/4$. We also report the results of the simulations if
stars are injected as a consequence of tidal binary disruption when $\eta=1$. As discussed, while the number density has the same slope of the Bahcall-Wolf solution for $r<r_{inj}$ (with a higher normalization), the cusp develops a steeper profile for $r_{inj}<r<\eta R$. The extent of the steeper profile depends both on the binary fraction near the SMBH influence radius, $\eta$, and on the fraction of injected stars that are not disrupted by the SMBH, $\omega<1$. As previously discussed (see Eq.~\ref{eqn:circbin}), a further $\ln(a_{\max}/a_{\min})$ factor has to be taken into account. As a consequence, the effective fraction to consider in Eq.~\ref{eqn:cuspbinaries} would be
\begin{equation}
\eta_{eff}=\frac{\omega}{\ln(a_{\max}/a_{\min})}\eta\ ,
\label{eqn:etaeff}
\end{equation}

In \citet{frasar18}, we discussed the fraction of the stars injected in the vicinity of the SMBH from dissolved binaries that would circularize even though the high eccentricity of its orbit. We considered a Brownian process governed by a continuous diffusion equation, and found that the ratio of fluxes upwards to circular orbits compared to that downward to tidal disruption orbits is
\begin{equation}
\omega={\mathcal{F}_{circ} \over \mathcal{F}_{disrupt}}={\ln(J_0/J_{LC}) \over  \ln(J_c/J_0) }\ ,
\end{equation}
where $J_0$ is the angular momentum of the injected star. For the parameters of our Galaxy, $\ln(J_0/J_{LC})\sim\sqrt{\alpha}\sim 2$ and $J_c/J_0 \sim \sqrt{r_{inj}/r_p} \sim (M/m)^{1/6}\sim 10$. The ratio is therefore $\omega\sim 0.3$, and $\eta_{eff}\sim 0.1 \eta$. As a consequence, $\sim 70$\% of the injected stars diffuse to TDE orbits, enhancing their rates. From Eq.~\ref{eqn:etaeff}, only $\sim 10\%$ of the injected stars are not disrupted and the cusp results steeper from $r_{min}\sim 250$ AU up to $R=0.1\eta R_h$. In Fig.~\ref{fig:cuspbin}, we show that the results of our simulation nicely follow Eq.~\ref{eqn:cuspbinaries}, with $\eta_{eff}= 0.1 \eta$.

In the previous Sections, we discussed the orbits that origin TDEs and MS-EMRIs. The stars that undergo TDEs come from the outskirt of the SMBH influence radius on very eccentric orbits ($1-e\sim 10^{-5}$). In this case, the dominant effect is the continuous two-body interactions with other stars, which scatters stars onto plunging orbits, thus disrupting them within one orbital period. The stars that end their lives as MS-EMRIs inspiral by losing gradually their energy due to GW emission within the region $T_{GW}<T_{2B}$, while keeping the pericenter roughly constant. For $\eta=0$, we find that the rates are $\Gamma_{TDE} \sim 3.3 \times 10^{-6}$ yr$^{-1}$ for TDEs and $\Gamma_{MS-EMRI} \sim 4 \times 10^{-8}$ yr$^{-1}$ for MS-EMRIs, in nice agreement with the analytical predictions of Eq.~\ref{eqn:ratiorert}. The ratio is $\Gamma_{MS-EMRI}/\Gamma_{TDE}\sim 1.2$ \%. 

Injection of dissolved binaries may modify the above picture. As shown in Fig.~\ref{fig:binaries}, disrupted binaries would leave injected stars both in the two-body driven region and in the GW driven region ($T_{GW}>T_{2B}$ and $T_{GW}<T_{2B}$, respectively), thus possibly enhancing both rates. Moreover, the injection of binaries may modify the distribution of stars \citep{frasar18}, thus shifting the $T_{GW}=T_{2B}$ boundary line. In general, we expect different TDE and MS-EMRI rates from the case with no injection of stars ($\eta=0$). Figure~\ref{fig:tdeemris_sim} illustrates the TDE and MS-EMRI rates as a function of the binary fraction $\eta$. While even for small $\eta$'s the MS-EMRI rate becomes larger than the case of no binaries, the TDE rate is significantly larger only when $\eta\gtrsim 0.5$, and becomes $\sim 2$ times larger than the case of no binary disruptions. While with no injection of binaries $\Gamma_{MS-EMRI}$ is just a few percent of $\Gamma_{TDE}$, the MS-EMRI rate becomes $\sim 10$ times larger for $\eta\sim 0.1$ and $\sim 1/5$ of the TDE rate if the fraction of binaries is significant ($\eta\gtrsim 0.5$).

\section{Discussions and Conclusions}
\label{sect:conclusions}

The recent big advance in instruments dedicated to the observation of our GC SMBH and the future GW LISA mission offer the unprecedented opportunity to test theoretical models of the densest environment in the Universe \citep{lisa17,grav2018a,grav2018b}. In galactic nuclei harbouring an SMBH, the short-term dynamics is dominated by the deep potential well of the SMBH, while two-body interactions shape the distribution of stars on longer timescales, modifying their energy and angular momentum. Stars revolving around the SMBH near the edge of its influence sphere may be scattered into extremely-eccentric orbits, which eventually end up with the tidal disruption of the star itself if it crosses the tidal disruption sphere. Moreover, stars sufficiently close to the SMBH also suffer from energy loss due to GW radiation emission. If the energy loss is rapid enough, these stars will gradually inspiral onto the SMBH on a Peters timescale, becoming luminous in GWs in the mHz frequency band \citep{alex17}.

In this paper, we have provided a simplified analytical treatment of the scattering processes in galactic stellar nuclei, assuming all stars have the same mass. We have discussed how the interplay between two-body relaxation and gravitational wave emission modifies the slope of the inner cusp within the SMBH sphere of influence. We have calculated the TDE rate of stars that are tidally disrupted by the SMBH and the rate of stars that end up their lives as MS-EMRI, disrupted by the SMBH due to the slow emission of GW radiation. We have found that typically the rate in the latter case is just a few percent of the former, thus implying a few MS-EMRI events for hundred TDEs.

We have also discussed the role of binary disruptions in the ecology of the galactic nuclei dynamics. In \citet{frasar18}, we showed that stars injected on highly-eccentric orbits in the vicinity of the SMBH due to Hills binary disruption may modify the shape of the density cusp, if the injection rate is large enough. The high eccentricity of these injected stars makes them of high relevance in the context of TDE and MS-EMRI events, since they can either be disrupted by the SMBH on plunging orbits or on gradually-circularized orbits due to GW emission. We have shown that the MS-EMRI rate can almost approach the TDE rate in case the binary fraction at the SMBH influence radius is close to unity.

We note that in our model we do not take into account the coherent torques between slowly precessing orbits, i.e. the resonant relaxation process \citep{kocs11,kocs15}. Such mechanisms are expected to be relevant for $r\sim 1000$ AU, until relativistic precession decouples the GW inspiral from the residual torques of the background stars. This may be the case of the S-stars, whose orbits may have evolved due to the resonant relaxation process \citep{perg2009,perg2010,gil17}. Nevertheless, \citet{baror16} showed that the inclusion of the resonant relaxation in the dynamical processes has little impact on the estimated rates of TDEs and EMRIs.

We have ignored the role of physical collisions. Close enough to the SMBH, the orbital velocities become larger than the typical escape velocity from the surface of a star ($v_{esc}\sim 600 \kms$ for $1\msun$ star), and collisions become more likely than scattering events \citep{alex17}. We generalize this analysis and take into account highly eccentric orbits which are the sources of dynamical events like TDEs and MS-EMRIs. We find the phase space where collisions are important, and plot that in Fig.~\ref{fig:tdeemris}. The likely outcome of a stellar collision depends, apart from the relative mass ratio of the colliding stars, on the ratio between their relative velocity to their surface escape speed \citep{benz1987,trac2007,gabu2010}. Due to the large velocity dispersion in galactic nuclei, the physical collision of two equal-mass $1\msun$ stars will likely eject half of the total, thus creating a blue straggler, in the outermost regions of the cusp. Closer to the SMBH, the collisions could lead to stellar destruction rather than to mergers since the kinetic energy in the colliding star exceeds the binding energy. Collisions have been shown to possibly play some role in the depletion of the red giant population \citep{dale2009,bar2010,sch2018}, but they may have also some importance in shaping the distribution of solar-mass stars \citet{sills2005,dale2006}, whose impact on the galactic nuclei economy deserves future attention.

Finally, we have not considered a mass function both for the stars in the cusp and the injected stars, but only a single-mass population of $1\ \mathrm{M}_{\odot}$ stars. As a consequence, we also do not account for mass segregation as due to dynamical friction. Objects of different masses have generally different slopes of the cusp density. The more massive the object the steeper the cusp profile due to more efficient segregation towards the center, with stellar black holes leading with the largest value $\rho_{BH}(r)\propto r^{-\alpha_{BH}}$, where $\alpha_{BH}\approx 1.5$-$2$ \citet{bahcall76,bahcall77,ale09,aharon16,baumg18}. Massive stellar remnants would also reduce the two-body relaxation time at small radii \citep{vasi2019}, thus changing the relative ratio of MS-EMRIs and TDEs (Eq.~\ref{eqn:ratiorert}) in favour of MS-EMRIs.  
We leave a detailed calculation to a future study. This also might effect the formation of binary stars and black holes that can merge as a consequence of the Kozai-Lidov effect \citep{ant12,step2016MNRAS,fragrish2018,grishin2018,hoa2018,fanto2019,flp2019M,steph2019}. All the picture is even more complicated by the possible presence of intermediate mass black holes and other remnants brought by inspiralling star clusters \citep{fgk18,flgk18}.

\section{Acknowledgements}

We thank Scott Tremaine and Brian Metzger for useful comments and discussions. RS is supported by an iCore and an ISF grant. GF is supported by the Foreign Postdoctoral Fellowship Program of the Israel Academy of Sciences and Humanities. GF also acknowledges support from an Arskin postdoctoral fellowship at the Hebrew University of Jerusalem.

\bibliographystyle{yahapj}
\bibliography{references}

\end{document}